\documentclass[useAMS,usenatbib]{mn2e}
\usepackage{txfonts}
\usepackage{natbib}
\usepackage{aas_macros}
\usepackage[all]{xy}

\usepackage[dvips]{graphicx}

\usepackage{hyperref}
\usepackage[usenames,dvipsnames]{color}
\definecolor{darkblue}{rgb}{0,0,.5}
\hypersetup{bookmarksopen,
			pdfstartview={FitH},
			colorlinks=true,
			breaklinks=true,
			linkcolor=darkblue,
			menucolor=darkblue,
			urlcolor=Red,
			citecolor=Green,
			linkcolor=blue}
\usepackage[all]{hypcap}

\def\del#1{{}}
\newcommand{\dd}{\mathrm{d}}
\newcommand{\eqref}[1]{(\ref{#1})}
\newcommand{\fNL}{f_{\mathrm{NL}}}
\newcommand{\ps}[1]{C^{\Theta\Theta}_{\ell_{#1}}}

\title[CMB lensing power reconstruction and primordial non-Gaussianity]{The interplay of CMB temperature lensing power reconstruction with primordial non-Gaussianity of local type}
\author[Philipp M. Merkel and Bj{\"o}rn Malte Sch\"afer]
{Philipp M. Merkel$^1$\thanks{e-mail: philipp.merkel@urz.uni-heidelberg.de} and Bj{\"o}rn Malte Sch\"afer$^2$\\
${}^1$Institut f{\"u}r Theoretische Astrophysik, Zentrum f{\"u}r Astronomie, Universit{\"a}t Heidelberg, Albert-Ueberle-Stra{\ss}e 2, 69120 Heidelberg, Germany\\
${}^2$Astronomisches Recheninstitut, Zentrum f{\"u}r Astronomie, Universit{\"a}t Heidelberg, M{\"o}nchhofstra{\ss}e 12, 69120 Heidelberg, Germany}

\begin{document}
\pagerange{\pageref{firstpage}--\pageref{lastpage}}
\pubyear{2010}
\maketitle
\label{firstpage}

% --- abstract --- %
\begin{abstract}
In the current era of high-precision CMB experiments, the imprint of gravitational lensing on the CMB temperature is exploited as a source of valuable information. Especially the reconstruction of the lensing potential power spectrum is of great interest. 
The reconstruction from the optimal quadratic estimator of the lensing potential, though, is biased. As long as the intrinsic CMB fluctuations are Gaussian this bias is well understood and controlled. In the presence of non-Gaussian primordial curvature perturbations, however, the CMB also acquires a non-Gaussian structure mimicking the lensing signal. Concentrating on primordial non-Gaussianity of local type, we address the resulting bias 
by extracting the lensing potential power spectrum from large samples of simulated lensed CMB temperature maps comprising different values of $\fNL$.
We find that the bias is considerably larger than previous analytical calculations suggested.
For current values of $\fNL$ and a sensitivity like that of the Planck mission, however, the bias is completely negligible on all but the largest angular scales. 
\end{abstract}

% --- keywords --- %

% --- introduction --- %
\section{Introduction}
\label{sec_introduction}

The cosmic microwave background (CMB) carries an incomparable content of cosmological information. After removing all foregrounds its tiny temperature fluctuations observed today mirrors the physical conditions in the Universe approximately 400,000 years after the big bang. The primary tool to access this information is the CMB temperature power spectrum. Additionally, CMB polarization is an important diagnostic of gravitational waves. Gravitational lensing of the CMB by the intervening large scale structure, however, alters the shape of the observed power spectra \citep[see][for a review]{2006PhR...429....1L}. In particular, it inevitably introduces $B$-modes confusing the imprint of primordial tensor perturbations. Therefore, in the context of parameter analysis and gravitational wave detection CMB lensing can be regarded as contaminant.

However, the unprecedented sensitivity of ongoing and future CMB experiments allows for a new perspective: The lensing signal in the CMB temperature itself contains valuable cosmological information. Precise knowledge of the lensed CMB power spectrum can be used to improve on parameter estimations from the primary CMB breaking several degeneracies \citep{2002PhRvD..65b3003H,1999MNRAS.302..735S,2006PhRvD..74l3002S}. Furthermore, lensing probes the (integrated) matter distribution from today's observer up to the last scattering surface. As the growth of large scale structure is sensitive to neutrino mass so is the lensing signal thereby tightening up the mass limits \citep{2006PhRvD..73d5021L}.
Besides these applications in cosmological parameter studies, direct reconstruction of the lensing effect is also of major importance. In order to access the primordial $B$-mode signal the lensing effect needs to be removed from the observed CMB requiring the lensing potential \citep{2002PhRvL..89a1303K,2004PhRvD..69d3005S}.

There are several approaches to extract the lensing potential from the observed CMB \citep{1999PhRvD..59l3507Z,2000PhRvD..62d3517G,2001ApJ...557L..79H,2003PhRvD..68h3002H,2003PhRvD..67h3002O}. \citet{2003PhRvD..67h3002O}  is very well suited for the instrumental noise level of current experiments regarding precision and numerical complexity. Being quadratic in the lensed temperature field this estimator exploits the lensing induced non-diagonal elements of the CMB covariance matrix. In the next step the reconstructed lensing potential can be used to estimate its power spectrum. The resulting estimator, however, is biased. 
There are several contributions to this bias, which can be classified according to the order of the lensing potential involved. The zeroth order contribution is dominant and given by the Gaussian variance of the estimator of the potential. Smaller but substantial contributions are of first and second order in the lensing potential and arise from the non-trivial part of the (lensing induced) CMB trispectrum.
Up to date all relevant contributions to this so-called reconstruction bias are identified and under control \citep{2003PhRvD..67l3507K,2011PhRvD..83d3005H}.

Since not only the information about the lensing potential power spectrum itself, but in part also the reconstruction bias, originates from the connected part of the \emph{lensing induced} CMB four-point function, any kurtosis present in the observed CMB will affect the power spectrum extraction, too. 
A non-trivial four-point function may arise from primordial non-Gaussian curvature perturbations, which are predicted by all inflationary models. Current observations suggest that the deviations from Gaussianity are quite small. Nevertheless, depending on the inflationary scenario, they can be rather sizable.

In this paper we extend the analytical work of \citet{2005PhRvD..71j3514L} to the full sky and investigate the impact of an intrinsic non-trivial CMB trispectrum on the lensing power spectrum reconstruction using numerical simulations.

The structure of this paper is the following: in Section~\ref{sec_power_reconstruction} we briefly review the key aspects of CMB lensing power extraction. Then, in Section~\ref{sec_local_non_gaussianity},  primordial non-Gaussian curvature perturbations of local type and their connection to a non-trivial CMB trispectrum are explained in detail. Our method to determine the additional reconstruction bias in the presence of primordial non-Gaussianity is presented along with the results in Section~\ref{sec_power_recon_bias}. Finally, Section~\ref{sec_summary} is devoted to a short summary of the main findings of this work.

As reference cosmology we choose the parameters compiled in the WMAP5+BAO+SN data set\footnote{The parameter values can be obtained from \url{http://lambda.gsfc.nasa.gov/product/map/dr3/parameters.cfm}.}. The matter content is described by $\Omega_c h^2= 0.1143$ and $\Omega_b h^2= 0.02256$, while the cosmological constant is given by $\Omega_\Lambda = 0.721$. Primordial perturbations are characterized by the scalar amplitude $A=2.457\cdot10^{-9}$ and spectral index $n_S=0.96$. The value of the Hubble constant measured today is $H_0 = 100\,h\,\mathrm{km}\,\mathrm{s}^{-1}\,\mathrm{Mpc}^{-1}$ with $ h=0.701$ and the optical depth is assumed to be $\tau = 0.084$.

% --- section: CMB lensing power spectrum reconstruction --- %
\section{CMB lensing power spectrum reconstruction}
\label{sec_power_reconstruction}

On their way from the last scattering surface to today's observer CMB photons are deflected by the intervening large scale structure. This gravitational lensing effect on the CMB is well described by a simple remapping of the temperature contrast, $\Theta(\mathbf{\hat{n}}) = \Delta T(\mathbf{\hat{n}}) / T_{\mathrm{CMB}}$,
 \begin{equation}
  \tilde\Theta (\mathbf{\hat{n}}) = \Theta \left[\mathbf{\hat{n}} + \nabla \phi(\mathbf{\hat{n}})\right]
 \end{equation}
 where a tilde denotes lensed quantities. The transition from the original direction on the sky to the lensed one is mediated by the angular gradient of the lensing potential $\phi(\mathbf{\hat{n}})$, i.e. the projected Newtonian gravitational potential $\Psi(\mathbf{\hat{n}})$
\begin{equation}
 \phi(\mathbf{\hat{n}}) = \frac{2}{c^2} \int_0^{\chi^*} \dd \chi \, \frac{\chi^*-\chi}{\chi^*\chi} \Psi(\chi \mathbf{\hat{n}}, \chi).
\end{equation}
This integral is formulated for a flat universe where the angular diameter distance coincides with the comoving distance $\chi$. It ranges to the last scattering surface at comoving distance $\chi^*$.
 
In harmonic space and linear in the deflection field the effect of lensing can be expressed as \citep{2000PhRvD..62d3007H}
\begin{equation}
\label{eq_lensing_effect}
 \tilde{\Theta}_{\ell m} \approx \Theta_{\ell m} + \sum_{LM\ell_1 m_1} (-1)^{m} \phi_{LM} \Theta_{\ell_1 m_1}
 \left(\begin{array}{ccc}
 \ell & L & \ell_1 \\
 -m  & M &  m_1
 \end{array}\right)F_{\ell L \ell_1}
\end{equation}
with
\begin{equation}
 F_{\ell_1 L \ell_2} = \frac{\Xi^2_L - \Xi^2_{\ell_1} + \Xi^2_{\ell_2}}{\sqrt{16\pi}} \Pi_{\ell_1 L \ell_2}
  \left(\begin{array}{ccc}
 \ell_1 & L & \ell_2 \\
 0  & 0 &  0
 \end{array}\right)
\end{equation}
where
\begin{equation}
 \Xi_{a\cdots n}\equiv \sqrt{a(a+1)\cdots n(n+1)}, \ \  \ \Pi_{a\cdots n} \equiv \sqrt{(2a+1)\cdots(2n+1)}.
\end{equation}
The sixfold indexed quantities in braces denote Wigner~3$j$ symbols \citep[see][]{1972hmfw.book.....A}.

\citet{2003PhRvD..67h3002O} realized that it is possible to extract the lensing potential from equation~\eqref{eq_lensing_effect} by taking the correlator of the lensed multipoles while keeping the realization of the lenses fixed. This procedure returns the multipoles of the lensing potential weighted by a function just depending on the intrinsic CMB power spectrum. Since lensing correlates temperature modes across a whole multipole band (whose width is given by the corresponding band power in the deflection field) one can construct an estimator of the lensing potential as weighted sum over lensed multipoles:
\begin{equation}
 \label{eq_optimal_qu_estimator}
 \hat\phi_{LM} = A_L \sum_{\ell_1 m_1\ell_2 m_2} (-1)^M g_{\ell_1\ell_2}(L) 
   \left(\begin{array}{ccc}
 \ell_1 & \ell_2 & L \\
 m_1  & m_2 &  -M
 \end{array}\right)
 \tilde{\Theta}_{\ell_1 m_1} \tilde{\Theta}_{\ell_2 m_2}.
\end{equation}
Here we have slightly changed our notation. From now on multipoles with a tilde denote the \emph{observed} temperature fluctuations containing the lensing signal, as well as all kinds of experimental noise.
The weighted sum needs an appropriate normalization given by
\begin{equation}
 A_L^{-1} = \frac{1}{\Pi^2_L} \sum_{\ell_1 \ell_2} f_{\ell_1 L \ell_2} g_{\ell_1 \ell_2}(L).
\end{equation}
Under the assumption that both the intrinsic temperature fluctuations and the lensing potential are homogeneous and isotropic Gaussian random fields, i.e. 
\begin{equation}
 \left\langle X_{\ell m} X^*_{\ell' m'}\right\rangle = C^{XX}_\ell \delta_{\ell\ell'}\delta_{mm'}, \quad X \in \left\lbrace \Theta, \phi\right\rbrace,
\end{equation}
\citet{2003PhRvD..67h3002O} derived optimal weights
\begin{equation}
  g_{\ell_1\ell_2}(L) = \frac{f_{\ell_1 L \ell_2}}{2C_{\ell_1,\mathrm{obs}}^{\Theta\Theta}C_{\ell_2,\mathrm{obs}}^{\Theta\Theta}}
  	= \frac{C^{\Theta\Theta}_{\ell_1} F_{\ell_2 L \ell_1}+ C^{\Theta\Theta}_{\ell_2} F_{\ell_1 L \ell_2}}
		{2C_{\ell_1,\mathrm{obs}}^{\Theta\Theta}C_{\ell_2,\mathrm{obs}}^{\Theta\Theta}} , 
\end{equation}
where optimal means that they minimize the estimator's Gaussian variance. The spectrum $C^{\Theta\Theta}_{\ell,\mathrm{obs}}$ denotes the observed one, i.e. it contains the lensing signal and all kinds of instrumental noise and foregrounds.
 
Given the estimated multipoles of the lensing potential field one can proceed to estimate the corresponding power spectrum via
\begin{equation}
  C^{\hat{\phi}\hat{\phi}}_L = \frac{1}{2L +1}\sum_{M}\hat{\phi}_{LM} \hat{\phi}^{*}_{LM}  .
\end{equation}
Accordingly, the expectation value of the reconstructed lensing potential power spectrum involves the four-point function of the lensed temperature contrast
\begin{eqnarray}
 \left\langle C^{\hat{\phi}\hat{\phi}}_L\right\rangle &=& \frac{A^2_L}{\Pi^2_L} \sum_{\ell_1m_1\ell_2m_2\atop\ell_3m_3\ell_4m_4} \sum_M
 (-1)^M
  \left(\begin{array}{ccc}
 \ell_1 & \ell_2 & L \\
 m_1  & m_2 &  -M
 \end{array}\right)\nonumber\\
 && \times
  \left(\begin{array}{ccc}
 \ell_3 & \ell_4 & L \nonumber\\
 m_3  & m_4 & M
 \end{array}\right)
 g_{\ell_1\ell_2}(L)g_{\ell_3\ell_4}(L)\\
 && \times \left\langle \tilde\Theta_{\ell_1m_1} \tilde\Theta_{\ell_2m_2} \tilde\Theta_{\ell_3m_3} \tilde\Theta_{\ell_4m_4} \right\rangle.
 \label{eq_expectation_value_of_estimator}
\end{eqnarray}
This estimator, however, is biased. While the connected part of the four-point function contains the information about the true lensing potential power spectrum, its Gaussian part gives the largest contribution to the bias which is identical with the normalization of the estimator, i.e. $N^{(0)}_L=A_L$. In addition, there are two more terms that need to be considered. The first one is linear in the lensing potential spectrum. It was first derived by \citet{2003PhRvD..67l3507K}. Since it is only substantial on small scales a treatment in the limit of a flat sky is well applicable
\begin{eqnarray}
 N_L^{(1)} &=& A^2(L) \int\frac{\dd^2 l_1}{(2\pi)^2} \int\frac{\dd^2 l_3}{(2\pi)^2} g(\mathbf l_1, \mathbf l_2) g(\mathbf l_3, \mathbf l_4) 
 	\nonumber\\
 && \times \left( C^{\phi\phi}_{|\mathbf l_1 - \mathbf l_3|} f(-\mathbf l_1, \mathbf l_3 ) f(-\mathbf l_2, \mathbf l_4 ) \right. \nonumber \\ 
 && + \left. C^{\phi\phi}_{|\mathbf l_1 - \mathbf l_4|} f(-\mathbf l_1, \mathbf l_4 ) f(-\mathbf l_2, \mathbf l_3 )\right)
 \label{eq_N_one}
\end{eqnarray}
with the additional constraint $\mathbf l_1 + \mathbf l_2 = \mathbf L = \mathbf l_3 + \mathbf l_4$.
The explicit expressions for the flat-sky analogues of the normalization and weights are given in Appendix~\ref{sec_flat_sky_expressions}.

\begin{figure}
\centering
\resizebox{\hsize}{!}{\includegraphics[]{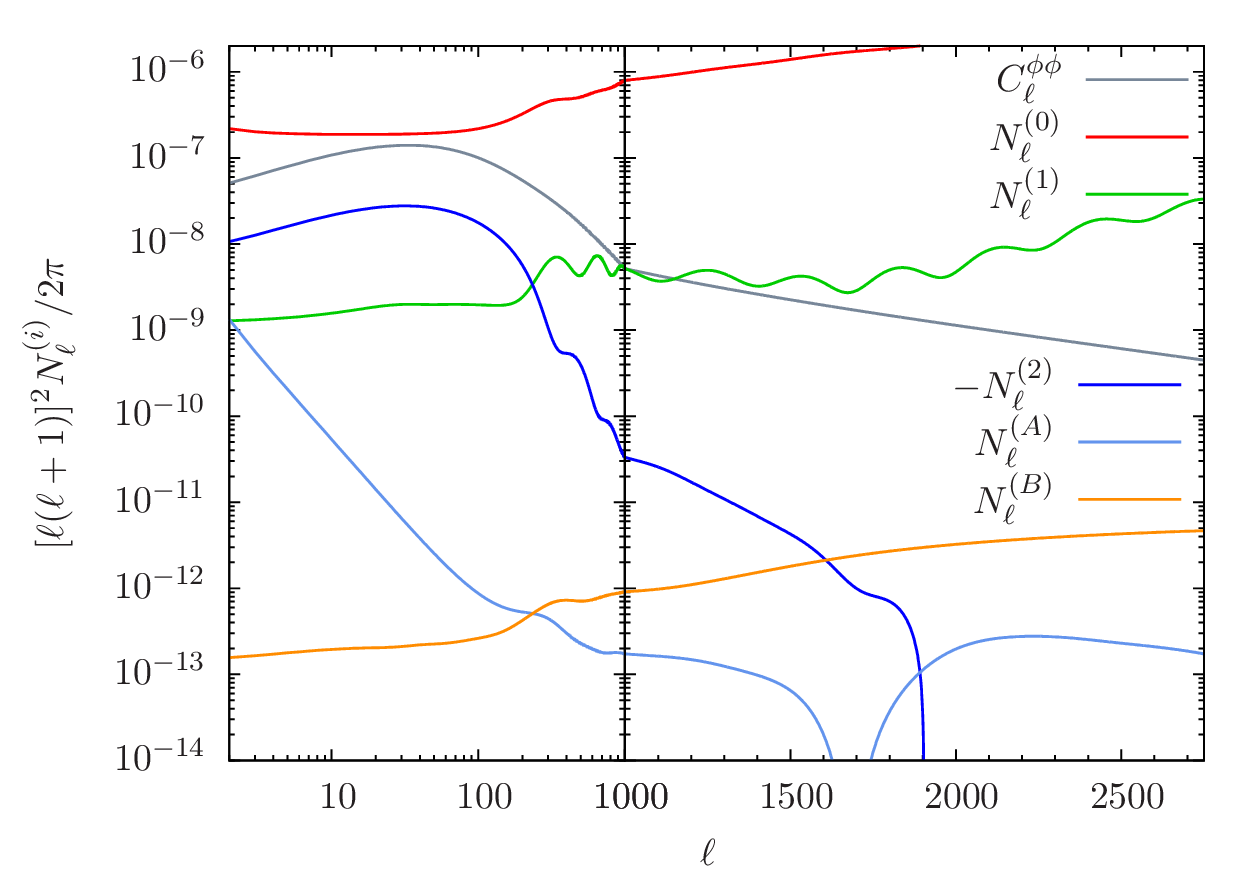}}
\caption{Biases involved in the reconstruction of the CMB lensing potential power spectrum $C^{\phi\phi}_\ell$, starting with the optimal quadratic estimator of the lensing potential defined in equation~\eqref{eq_optimal_qu_estimator}. The terms $N^{(A)}_\ell$ and $N^{(B)}_\ell$ are analytical estimates of the additional bias arising from primordial non-Gaussian curvature perturbations of local type where $\fNL=100$ has been assumed.}
\label{fig_all_biases}
\end{figure}

Furthermore on large scales \citet{2011PhRvD..83d3005H} identified a negative bias which is quadratic in the lensing potential spectrum. Its approximative formula
\begin{eqnarray}
 N_L^{(2)} &\approx& 4 C^{\phi\phi}_L \left( \frac{A_L}{\Pi^2_L}\right)^2 \left( \sum_{\ell_3\ell_4} g_{\ell_3\ell_4}(L)f_{\ell_3L\ell_4}\right) \nonumber \\
 && \times \sum_{\ell_1\ell_2} g_{\ell_1\ell_2}(L) F_{\ell_2L\ell_1} \left( C^{\tilde{\Theta}\tilde{\Theta}}_{\ell_1} - C^{\Theta\Theta}_{\ell_1}\right)
\end{eqnarray}
is in excellent agreement with results from numerical simulations.
All three biases are shown in Figure~\ref{fig_all_biases}.

% --- section: Local Non-Gaussianity --- %
\section{Local Non-Gaussianity}
\label{sec_local_non_gaussianity}

In the derivation of the optimal reconstruction weights and the three different biases of the estimated lensing potential power spectrum the intrinsic CMB fluctuations were always assumed to be Gaussian distributed. 
In this case, and under idealized experimental conditions, the non-Gaussian structure of the observed CMB solely arises from the lensing effect. Non-Gaussianity of the lensed CMB then manifests itself in the connected parts of the $n$-point correlation functions. Odd moments are not generated by lensing as long as both fields are uncorrelated. Correlations, however, do arise on large scales due to the integrated Sachs-Wolfe (ISW) effect: the cosmic large-scale structure does not only deflect the CMB photons but its late-time evolution also gives rise to an additional (large-scale) temperature anisotropy. On small scales there are correlations because of the non-linear ISW (or Rees-Sciama) effect, as well as the Sunyaev-Zel'dovich effects. These correlations result in a non-vanishing bispectrum of the lensed CMB. Thus, there are non-Gaussian signatures present in the lensed, i.e. observed, CMB; even when starting from Gaussian distributed temperature fluctuations.
 
Inflationary models, however, predict at least small deviations from pure Gaussian initial conditions. While simple inflationary scenarios generate a nearly Gaussian spectrum of the primordial curvature perturbations \citep{1981PhRvD..23..347G,1983PhRvD..28..679B,1992PhR...215..203M}, more involved models are expected to introduce a significant amount of non-Gaussianity. These models can be classified by the shape of their bispectrum: besides \emph{equilateral} \citep{2004PhRvD..70l3505A,2004JCAP...04..001A,2004PhRvD..70j3505S,2005JCAP...06..003S,2007JCAP...01..002C,2008JHEP...03..014C,2008JCAP...03..028L} and \emph{orthogonal} type \citep{2010JCAP...01..028S}, \emph{local} non-Gaussianity is an important category \citep{2003PhRvD..67b3503L,2004PhRvD..69b3505D,2008JCAP...06..030B,2008JCAP...10..008B,2010JCAP...11..037A,2011JCAP...11..005E}.

Primordial non-Gaussian curvature perturbations of local type can be described starting from a Gaussian fluctuation field $\Phi_\mathrm{L}(\mathbf x)$ and setting
\begin{equation}
 \label{eq_def_phi_nl}
 \Phi_{\mathrm{NL}}(\mathbf x ) = \Phi_\mathrm{L}(\mathbf x ) + 
 \fNL \left( \Phi_\mathrm{L}^2(\mathbf x ) - \left\langle\Phi_\mathrm{L}^2(\mathbf x )\right\rangle \right).
\end{equation}
The amplitude of the non-Gaussianity observed is then set by the dimensionless parameter $\fNL$ which in this case is assumed to be constant although models with scale dependent amplitude do exist \citep{2010JCAP...10..004B}.

The ansatz of equation~\eqref{eq_def_phi_nl} can be extended by adding a further term of the form $g_{\mathrm{NL}} \Phi^3_\mathrm{L}(\mathbf x)$. We ignore this term in our analysis for simplicity and because $g_{\mathrm{NL}}$ is observationally almost completely unconstrained. However, in order to be competitive with the $\fNL$ contributions, its value must be of the order $\sqrt{g_{\mathrm{NL}}}\sim\fNL$ but is not constrained by theory, in fact it is possible to tune the inflationary model arbitrarily in these parameters \citep{2006PhRvD..74j3003S}. Likewise, terms proportional to $\tau_{\mathrm{NL}}$, which are similar in magnitude relative to $f_{\mathrm{NL}}$ because of the Suyama-Yamaguchi relation $\tau_{\mathrm{NL}}\geq(6/5\:f_{\mathrm{NL}})^2$ \citep[see, e.g.][]{2011PhRvL.107s1301S, PhysRevD.77.023505, PhysRevLett.106.251301, 1475-7516-2010-12-030} are neglected. Neglecting these contributions, however, should at most result in a slight underestimation of the impact of local primordial non-Gaussianity on the lensing power reconstruction.

By construction, the four-point function of the curvature perturbations acquires a non-trivial part. This is also true for the intrinsic CMB fluctuations since they are simply related to those in the curvature field via an integration over the corresponding radiation transfer function $\Delta_\ell(k)$
\begin{equation}
 \Theta_{\ell m} = 4\pi \,(-\mathrm{i})^\ell \int \frac{\dd^3 k}{(2\pi)^3} \; \Phi_{\mathrm{NL}}(\mathbf k) \Delta_\ell(k) Y^*_{\ell m} (\mathbf{\hat{n}}_k)
\end{equation}
with $\Phi_{\mathrm{NL}}(\mathbf k)$ being the Fourier transform of equation~\eqref{eq_def_phi_nl}.

The non-trivial part of the angular trispectrum of the CMB temperature contrast is most conveniently expressed in its fully reduced form (at leading order in $\fNL$)
\begin{eqnarray}
  \mathcal{T}^{\ell_1\ell_2}_{\ell_3\ell_4}(L) &=& \int r_1^2 \dd r_1 r_2^2 \dd r_2 F_L(r_1,r_2)\alpha_{\ell_1} ( r_1) \beta_{\ell_2}(r_1)
  				\nonumber\\
 			& &\phantom{\int} \times \alpha_{\ell_3} ( r_2) \beta_{\ell_4}(r_2) h_{\ell_1 L \ell_2}  h_{\ell_3 L \ell_4}
			\label{eq_fully_red_tri_spec}
\end{eqnarray}
with
\begin{equation}
 \label{eq_cmb_trispectrum_F}
 F_L(r_1,r_2) = \frac{2}{\pi} \int k^2 \dd k\;P_{\Phi\Phi} ( k ) j_L(kr_1) j_L(kr_2),
\end{equation}
\begin{equation}
 \label{eq_cmb_trispectrum_alpha}
 \alpha_\ell(r) = 2\fNL\frac{2}{\pi} \int k^2 \dd k\;  \Delta_\ell(k) j_\ell(kr),
\end{equation}
\begin{equation}
 \label{eq_cmb_trispectrum_beta}
 \beta_\ell(r) = \frac{2}{\pi} \int k^2 \dd k\;  P_{\Phi\Phi} ( k )\Delta_\ell(k) j_\ell(kr)
\end{equation}
and
\begin{equation}
 h_{\ell_1 L \ell_2} = \frac{\Pi_{\ell_1 L \ell_2}}{\sqrt{4\pi}}
 \left(\begin{array}{ccc}
 \ell_1 & L & \ell_2 \\
 0  & 0 &  0
 \end{array}\right)
\end{equation}
\citep[see][for details on this formalism]{2002PhRvD..66f3008O}. In the expressions above, $P_{\Phi\Phi}(k)$ is the power spectrum of the curvature perturbations and $j_\ell(r)$ denotes the $\ell$-th spherical Bessel function of the first kind \citep[see][]{1972hmfw.book.....A}.

The computation of the trispectrum given in equation~\eqref{eq_fully_red_tri_spec} is numerically quite cumbersome. \citet{2002PhRvD..66f3008O}, however, found a very useful approximation formula by extending the Sachs-Wolfe approximation, i.e.
\begin{equation}
 \Delta_\ell(k) = \frac{1}{3} j_\ell\left[k (\eta_0 - \eta_{\mathrm{rec}})\right],
\end{equation}
valid at small $\ell$ into the acoustic regime
\begin{equation}
 \label{eq_trispectrum_approx}
 \mathcal{T}^{\ell_1\ell_2}_{\ell_3\ell_4}(L) \approx 36 h_{\ell_1 L \ell_2} h_{\ell_3 L \ell_4}\fNL^2 C_L^{\mathrm{SW}} \ps{2}\ps{4}
\end{equation}
with
\begin{equation}
 C_L^{\mathrm{SW}} = \frac{2}{9\pi} \int k^2 \dd k \; P_{\Phi\Phi}(k) j^2_L\left[k (\eta_0 - \eta_{\mathrm{rec}})\right]
\end{equation}
where $\eta_0 - \eta_{\mathrm{rec}}$ is the conformal time spread between today and recombination.
This approximation holds as long as the last scattering surface can be considered as thin and its temperature fluctuations only vary slowly.

% --- section: Power reconstruction bias --- %
\section{Power reconstruction bias}
\label{sec_power_recon_bias}
 
\subsection{Zeroth order calculation}

Since the first and second order bias of the reconstructed lensing potential power spectrum arise from the connected part of the (lensing induced) CMB trispectrum, one expects that also an intrinsic (non-trivial) trispectrum of the unlensed CMB contributes an additional bias. This bias was first addressed by \citet{2005PhRvD..71j3514L}. In their derivation they assumed that both of the effects leading to a connected part of the observed CMB four-point function are completely separable. Thus, they did not consider a so-to-speak lensed trispectrum but just added the connected part of the unlensed CMB trispectrum. In this approximation, and in the limit of a flat sky, the additional bias is given by
\begin{eqnarray}
   N^{(\fNL,0)}(L) &=& A^2(L) \int\frac{\dd^2 l_1}{(2\pi)^2}\int\frac{\dd^2 l_3}{(2\pi)^2}g(\mathbf l_1, \mathbf l_2) g(\mathbf l_3, \mathbf l_4)
  				\nonumber\\
			& & \times \left(\mathcal{P}^{l_1l_2}_{l_3l_4}(L) + 2 \mathcal{P}^{l_1l_3}_{l_2l_4}(|\mathbf l_1 - \mathbf l_3|)\right)
  \label{eq_analytic_flat_sky_zeroth_bias}
\end{eqnarray}
with the additional constraints $\mathbf l_1 + \mathbf l_2 = \mathbf L$ and $\mathbf l_3 + \mathbf l_4 = \mathbf L$.
The functional form of the flat-sky expressions are relegated to Appendix~\ref{sec_flat_sky_expressions}.
\citet{2005PhRvD..71j3514L} also computed the bias contribution from the $g_{\mathrm{NL}}$ term which is not considered in this work.

Starting from equation~\eqref{eq_expectation_value_of_estimator}, the all-sky generalization of equation~\eqref{eq_analytic_flat_sky_zeroth_bias} is
\begin{eqnarray}
 N^{(\fNL,0)}_L &=& \left(\frac{4A_L^2}{\Pi^4_L}\right) \sum_{\ell_1\ell_2\ell_3\ell_4} g_{\ell_1\ell_2}(L) g_{\ell_3\ell_4}(L) \nonumber\\
 			&& \times \biggl[ \mathcal{T}^{\ell_1\ell_2}_{\ell_3\ell_4}(L) + \Pi_L^2 \sum_{L'}
			\left\lbrace\begin{array}{ccc}
			\ell_1 & \ell_2 & L \\
			\ell_4 & \ell_3 & L'
			\end{array}\right\rbrace
			\biggr. \nonumber\\
			&& \biggl. \times \left( (-1)^{\ell_2 + \ell_3} \mathcal{T}^{\ell_1\ell_3}_{\ell_2\ell_4}(L')
			+(-1)^{L + L'} \mathcal{T}^{\ell_1\ell_3}_{\ell_4\ell_2}(L') \right) \biggr].
			\label{eq_analytic_all_sky_zeroth_bias}
\end{eqnarray}
The quantity in curly brackets denotes a Wigner~$6j$ symbol \citep[see][]{1972hmfw.book.....A}.
Using the approximative formula for the trispectrum of unlensed temperature fluctuations given in  equation~\eqref{eq_trispectrum_approx} the similarity with the lensing induced CMB trispectrum \citep{2001PhRvD..64h3005H}
\begin{equation}
 \mathbb{T}^{\ell_1\ell_2}_{\ell_3\ell_4}(L) = C^{\phi\phi}_L C^{\Theta\Theta}_{\ell_2} C^{\Theta\Theta}_{\ell_4} F_{\ell_1 L \ell_2} F_{\ell_3 L \ell_4}
\end{equation}
is striking. There are two important consequences. First, the term involving the Wigner-6$j$ symbol, hereafter denoted by $N^{(B)}_L$, should only contribute on small scales, so it can safely be treated in the flat sky-limit, i.e. by using the second part of equation~\eqref{eq_analytic_flat_sky_zeroth_bias} . Second, the other term, afterwards denoted by $N^{(A)}_L$, should be proportional to $C^{\mathrm{SW}}_L$ implying a scaling of the form
\begin{equation}
 N^{(A)}_L \propto \frac{1}{L^3(L+1)^3}
\end{equation}
due to the fact, that 
\begin{equation}
 C_L^{\mathrm{SW}} = \frac{2}{9\pi} \int k^2 \dd k \; P_{\Phi\Phi}(k) j^2_L\left[k (\eta_0 - \eta_{\mathrm{rec}})\right] \propto \frac{1}{L(L+1)}
\end{equation}
for a scale-invariant power spectrum \citep[cf.][]{2000cils.book.....L}.

In Figure~\ref{fig_all_biases} the two additional bias contributions $N^{(A)}_L$ and $N^{(B)}_L$ are shown. The agreement with the results of \citet{2005PhRvD..71j3514L} is very good although they used the exact temperature transfer functions instead of the extended Sachs-Wolfe approximation applied in this work. Only in the acoustic regime the $N^{(A)}_L$ term shows substantial differences. There, however, the bias is completely dominated by the $N^{(B)}_L$ term, which is accurately described by the extended Sachs-Wolfe approximation even on smallest scales as the comparison with Figure~1 of \citet{2005PhRvD..71j3514L} shows. On large scales, the full-sky calculation does not greatly improve the results of \citet{2005PhRvD..71j3514L} obtained neglecting the curvature of the sky. For both methods one recovers the expected scaling relation with the multipole order \citep[cf. again Figure~1 of][]{2005PhRvD..71j3514L}.

As already pointed out by \citet{2005PhRvD..71j3514L} for current values of $\fNL$, $ -10\leq \fNL \leq 74$ \citep{2011ApJS..192...18K}, the additional bias is not a severe contaminant of the reconstructed power spectrum. On almost all scales the bias is more than four orders of magnitude smaller than the biases arising from the lensing induced trispectrum. Only on largest scales the contributions from the $N^{(A)}_L$ term can be substantial. Consequently, as long as a fully separate perturbative treatment of both, gravitational lensing, as well as primordial non-Gaussianity, is applicable, the additional bias is well controlled and for current $\fNL$ parameters negligible.

However, the validity of this separation ansatz is not obvious. In the presence of primordial non-Gaussianity the mode coupling due to lensing is more involved since here the perturbative expansions of the lensed temperature field in powers of the lensing potential on the one hand and the unlensed temperature field in powers of the non-Gaussian curvature perturbations on the other hand compete. In order to circumvent the assumptions and limitations which are necessary for an analytical study of this problem we will resort to numerical simulations. Here all the different effects involved, primordial non-Gaussianity, lensing, instrumental noise, can be incorporated in a consistent manner.

\subsection{Numerical simulations}
\label{subsec_numerical_sim}

The starting point of the simulations is the large sample of non-Gaussian CMB realizations provided by \citet{2009ApJS..184..264E}. The non-Gaussian structure arises from primordial curvature perturbations of local type detailed in Section~\ref{sec_local_non_gaussianity}. The maps of \citet{2009ApJS..184..264E} are well suited for our analysis because they did not focus on the intrinsic CMB bispectrum but produced simulations with accurate non-Gaussian statistics to all correlation orders. Every realization consists of a set of Gaussian $(\Theta^{\mathrm{L}}_{\ell m})$ and non-Gaussian $(\Theta^{\mathrm{NL}}_{\ell m})$ multipoles, so that a CMB map containing the desired level of non-Gaussianity characterized by $\fNL$ can be built via
 \begin{equation}
 \label{eq_build_theta}
  \Theta_{\ell m} = \Theta^{\mathrm{L}}_{\ell m} + \fNL \Theta^{\mathrm{NL}}_{\ell m}.
 \end{equation}
The cosmological background of these simulations is adapted from the WMAP5+BAO+SN data as summarized at the end of 
Section~\ref{sec_introduction}.

Having a non-Gaussian CMB map at hand, this can be lensed in the framework of \textsc{LensPix}\footnote{\url{http://cosmologist.info/lenspix/}} 
\citep{2005PhRvD..71h3008L}. The remapping technique implemented in \textsc{LensPix} assumes the lensing potential being fully described by its power spectrum. Obviously, this is no longer true in the presence of primordial non-Gaussianity. However, for the reconstruction scheme detailed in Section~\ref{sec_power_reconstruction} the CMB and the lensing potential are treated as statistically independent fields and higher order statistics of the lensing potential are not involved.
 
Since the map is already given in multipole space we use a slightly modified version of \textsc{LensPix} which skips the simulation of a (Gaussian) CMB realization and starts directly from the temperature multipoles constructed according to equation~\eqref{eq_build_theta}. In order to get a lensed CMB map at \textsc{HEALPix} \citep{2005ApJ...622..759G} resolution $N_{\mathrm{side}} = 2048$ which is cosmically accurate up to $\ell_{\mathrm{max}} = 2750$ one needs input multipoles up to $\ell_{\mathrm{input}} = 3000$. The non-Gaussian maps of \citet{2009ApJS..184..264E}, however, cover only a multipole range up to $\ell_{\fNL} = 1024$. We therefore extend the maps in the range $\ell_{\fNL} < \ell \leq 3000 $ with pure Gaussian multipoles. Thus, equation~\eqref{eq_build_theta} needs to be replaced by
 \begin{equation}
 \label{eq_build_theta_mod}
  \Theta_{\ell m} = \left\lbrace\begin{array}{ll}
  \Theta^{\mathrm{L}}_{\ell m} + \fNL \Theta^{\mathrm{NL}}_{\ell m}, & \mathrm{if} \ \ell \leq 1024 \\
  \Theta^{\mathrm{L}}_{\ell m}, & \mathrm{else.}
  \end{array}
  \right.
 \end{equation}
The power spectrum needed for the construction of Gaussian CMB realizations has been computed using \textsc{CAMB}\footnote{\url{http://camb.info/}} \citep{2000ApJ...538..473L}.

This method is not expected to crucially influence the main conclusion of this work. The non-Gaussianity of the maps constructed according to this prescription will be even milder due to the fact that scales smaller than about ten arcminutes are not affected by the primordial non-Gaussian curvature perturbations. Likewise, if there are serious concerns about this method, one can just think of the resulting maps as not being subjected to primordial non-Gaussianity of the local type but subjected to some intermediate type which is scale dependent. In any case, the influence of the resulting non-trivial intrinsic CMB trispectrum on the reconstructed lensing power needs to be studied.
 
Finally, the lensed maps are exposed to instrumental noise only. But neither any further secondary anisotropies nor foregrounds nor experimental complications like incomplete sky coverage are included. Adapting the specification of the Planck satellite mission, the maps are first convolved with a Gaussian beam with $\sigma_{\mathrm{FWHM}} = 7\, \mathrm{arcmin}$. Before deconvolution, white Gaussian pixel noise with standard deviation $\sigma_N = 27\, \mu\mathrm{K}\,\mathrm{arcmin}$ is added. The power spectra of the final maps are well described by the analytic expression derived by \citet{1995PhRvD..52.4307K}
\begin{equation}
 C^{\Theta\Theta}_{\ell,\mathrm{obs}} = C^{\tilde\Theta\tilde\Theta}_\ell + \left( \frac{\sigma_N}{T_{\mathrm{CMB}}}\right)^2 \mathrm{e}^{\ell(\ell+1)\sigma^2_\mathrm{FWHM}/8\log 2} .
\end{equation}

\subsection{Efficient estimator}
\label{subsec_efficient_estimator}

For practical purposes, the estimator in the form of equation~\eqref{eq_optimal_qu_estimator} is not suitable. One should use its angular space representation instead
\begin{equation}
 \hat{\phi}_{LM} = A_L \int \dd\Omega \, Y^*_{LM}(\mathbf{\hat n})\nabla^i \left[ V(\mathbf{\hat n}) \nabla_i U (\mathbf{\hat n}) \right] .
\end{equation}
Here the spin-gradient derivative $\nabla$ just acts on the spin part of the corresponding fields, the inverse weighted temperature map
\begin{equation}
 V(\mathbf{\hat n}) = \sum_{\ell m} \frac{1}{C^{\Theta\Theta}_{\ell,\mathrm{obs}}} \tilde{\Theta}_{\ell m} Y_{\ell m}(\mathbf{\hat n})
\end{equation}
and the Wiener reconstruction of the unlensed temperature field
\begin{equation}
 U(\mathbf{\hat n}) = \sum_{\ell m} \frac{C^{\Theta\Theta}_\ell}{C^{\Theta\Theta}_{\ell,\mathrm{obs}}} \tilde{\Theta}_{\ell m} Y_{\ell m}(\mathbf{\hat n})
\end{equation}
\citep{2003PhRvD..67h3002O}.
This representation allows for the use of fast harmonic transform algorithms on the sphere.

\subsection{Results}

\begin{figure}
\centering
\resizebox{\hsize}{!}{\includegraphics[]{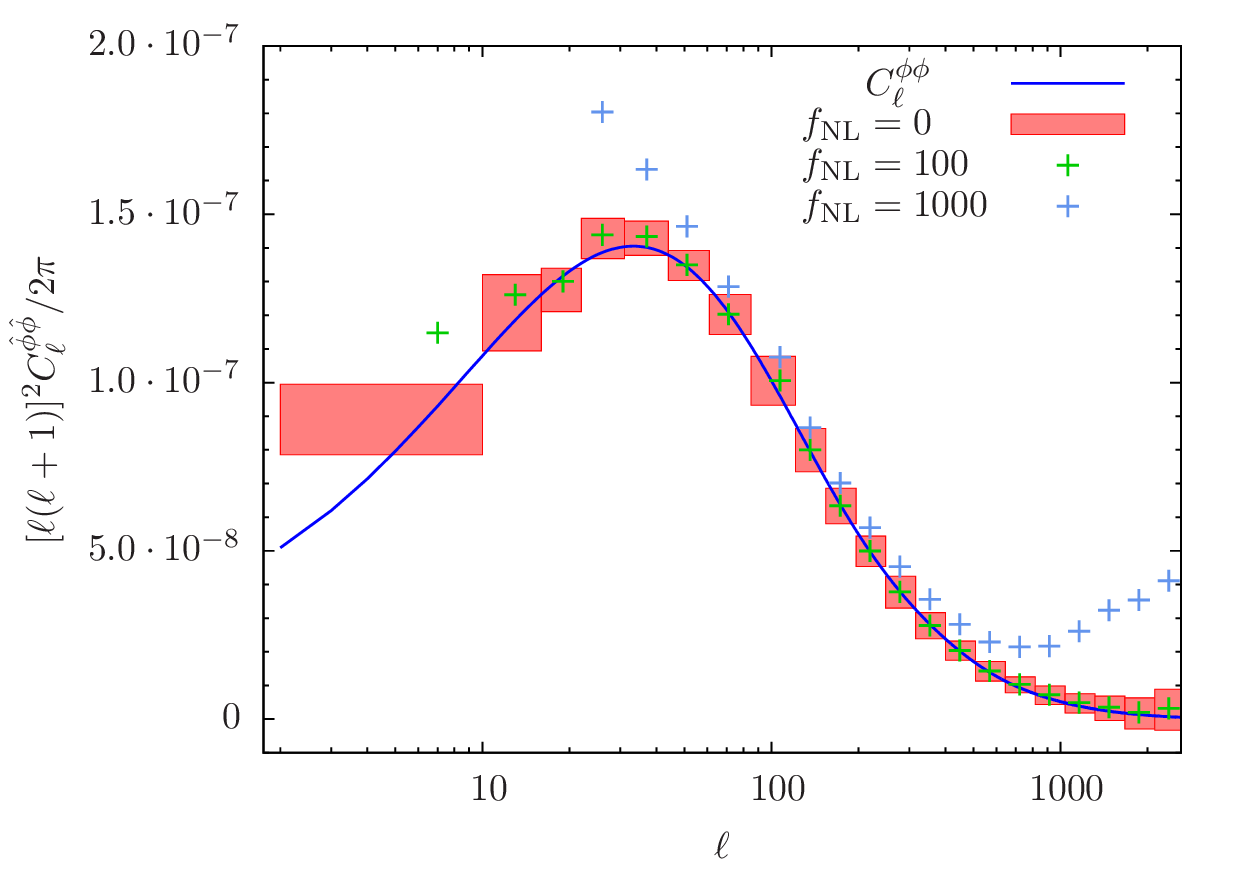}}
\caption{Illustration of the performance of the CMB lensing power reconstruction discussed in the text and condensed in 
equation~\eqref{eq_pow_spec_estimation_minus_biases}. The red boxes indicate the width of the multipole band over which the extracted power spectrum has been binned, as well as the $1\sigma$ error for one realization. In addition, the reconstructions in presence of primordial non-Gaussianity for two different values of $\fNL$ are shown. Note that on large angular scales the bias for $\fNL=1000$ exceeds the range of the plot by far (cf. Figure~\ref{fig_bias_fNL_1000}) and is not shown for clarity.}
\label{fig_example_of_recon}
\end{figure}
In order to determine the bias associated with the connected part of the trispectrum of the unlensed CMB which is generated by primordial non-Gaussian curvature perturbations of local type we proceeded as follows
 \begin{enumerate}
  \item[(1)] construct $\mathcal{N}$ CMB maps following equation~\eqref{eq_build_theta_mod} setting $\fNL=0$
  \item[(2)] use these maps to build a sample of lensed maps which contain the desired instrumental noise properties as 
 	detailed in Section~\ref{subsec_numerical_sim}
  \item[(3)] apply the efficient estimator of Section~\ref{subsec_efficient_estimator} to the sample and reconstructing the lensing potential power spectrum
 	for each map according to
 	\begin{equation}
		\label{eq_pow_spec_estimation_minus_biases}
  		C^{\hat{\phi}\hat{\phi},(i)}_L = \frac{1}{2L +1}\sum_{M}\hat{\phi}^{(i)}_{LM} \hat{\phi}^{*(i)}_{LM} - N^{(0)}_L - N^{(1)}_L - N^{(2)}_L
 	\end{equation}
	where the upper index runs from 1 to $\mathcal N$. The three bias contributions in the equation above have not been computed separately for 
	each individual realization. Instead, the theoretical power spectrum of the lensed and unlensed CMB, as well as of the lensing potential, again 
	computed with \textsc{CAMB}, were used.
	To illustrate that in this way the (purely lensing induced) biases are very well captured, we plot in Figure~\ref{fig_example_of_recon} the lensing potential power
	reconstruction for $\fNL=0$ obtained from hundred CMB realizations. The boxes indicate the width of the multipole band and the one sigma error
	for one realization.
  \item[(4)] compute the sample mean
 	\begin{equation}
 	\bar{C}^{\hat{\phi}\hat{\phi}}_L = \frac{1}{\mathcal{N}} \sum_{i=1}^\mathcal{N} C^{\hat{\phi}\hat{\phi},(i)}_L 
 	\end{equation}
  \item[(5)] repeat steps (1) to (4) but now setting $\fNL = 100$ in the first step
  \item[(6)] repeat step (5) but now starting with $\fNL=1000$
  \item[(7)] compute the bias for both values of $\fNL$ according to
  \begin{equation}
  \label{eq_extract_NfNL_bias}
   N_L^{(\fNL)} = \left.\bar{C}^{\hat{\phi}\hat{\phi}}_L\right|_{\fNL} - \left.\bar{C}^{\hat{\phi}\hat{\phi}}_L\right|_{\fNL=0}
  \end{equation}
 \end{enumerate}
It needs to be stressed that in steps (1) to (3) for all three values of $\fNL$ the simulation conditions have been the same. In other words, we kept track of the different random seeds involved in the map making procedure. Thus, the $i$-th map is always lensed by the same realization of the lensing potential and always contains the same realization of instrumental noise. This ensures that the differences in the reconstructed power spectra, i.e. the bias, are solely caused by the variations in $\fNL$. The numerical analysis of this work has been carried out setting $\mathcal{N}=100$.

\begin{figure}
\centering
\resizebox{\hsize}{!}{\includegraphics[]{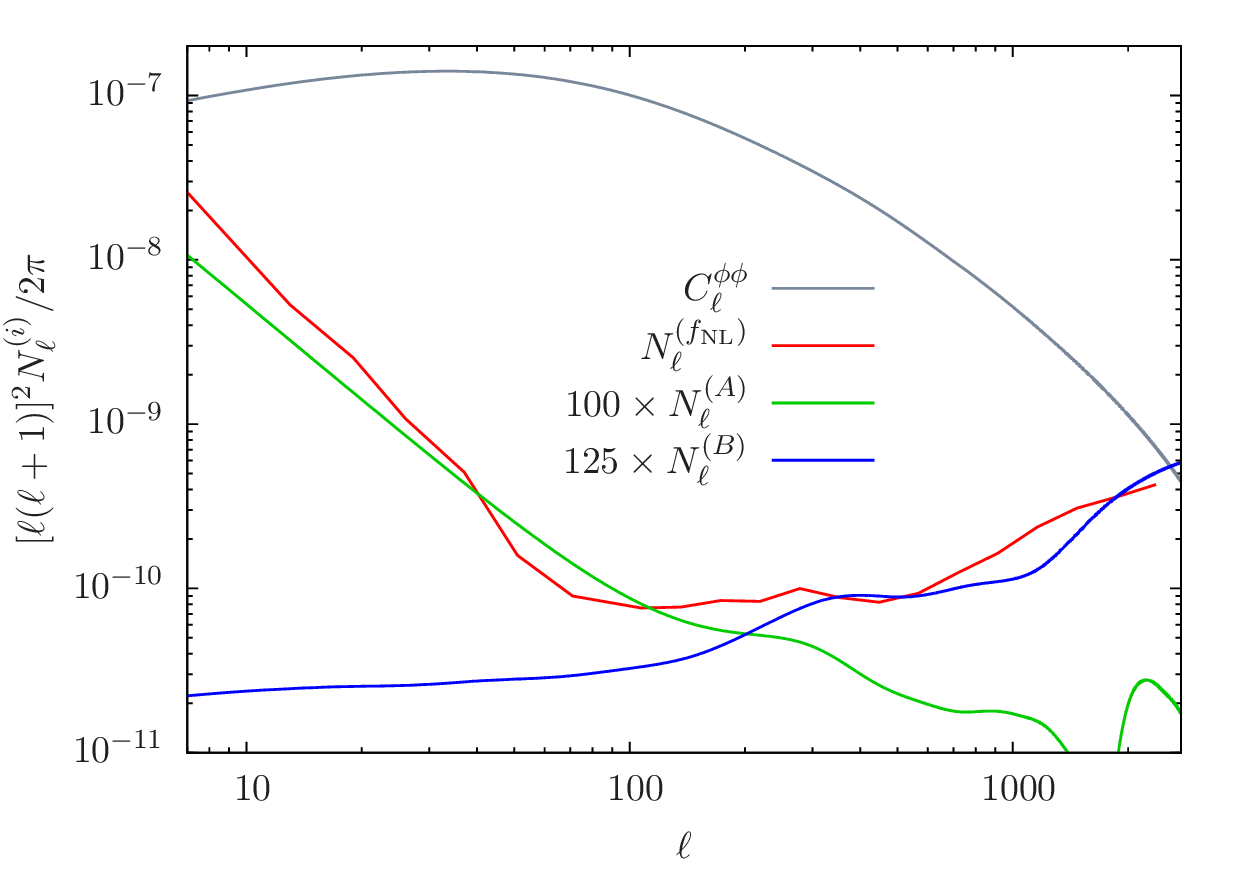}}
\caption{Additional bias associated with the presence of primordial non-Gaussian curvature perturbations. The bias (red curve) has been extracted from numerical simulations according to equation~\eqref{eq_extract_NfNL_bias} with $\fNL=100$. In addition, the analytical estimates of the bias (green and blue curve), phenomenologically amplified, are shown.}
\label{fig_bias_fNL_100}
\end{figure}

\begin{figure}
\centering
\resizebox{\hsize}{!}{\includegraphics[]{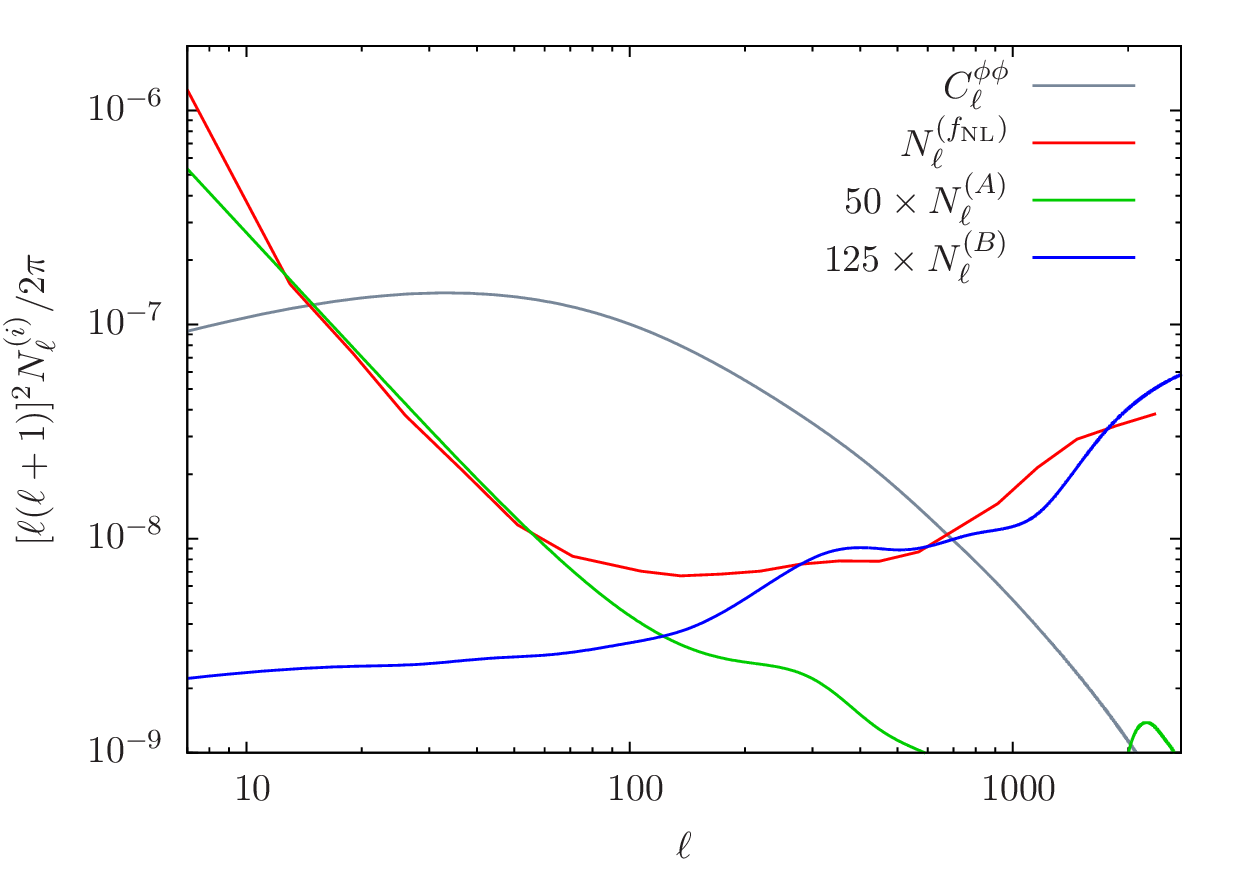}}
\caption{Same as Figure~\ref{fig_bias_fNL_100} but for another choice of the $\fNL$ parameter, namely $\fNL=1000$. Note that in this plot the $N^{(A)}_\ell$ term has been scaled differently highlighting that the bias found in numerical simulations does not exhibit a simple scaling relation with $\fNL$ in contrast to the analytical model.}
\label{fig_bias_fNL_1000}
\end{figure}

In Figure~\ref{fig_example_of_recon} the reconstructed power spectra for both values of $\fNL$ are plotted, while Figure~\ref{fig_bias_fNL_100} and Figure~\ref{fig_bias_fNL_1000} shows the corresponding biases. 
For comparison, also the bias derived by \citet{2005PhRvD..71j3514L} (cf. equation~\ref{eq_analytic_flat_sky_zeroth_bias} and~\ref{eq_analytic_all_sky_zeroth_bias}) is plotted. While the functional form of the bias found in the simulations resembles that of the analytical zero-th order prediction, its amplitude is much larger. Interestingly, the disagreement in amplitude is different for the two parts of the analytic bias formula. For the smaller value of $\fNL$ the $N^{(A)}_L$ part which gives the relevant contribution on large scales, needs to be scaled by a factor of roughly 100 to reach the same level as the bias found in the simulations. Its small scale counterpart, the $N^{(B)}_L$ term differs approximately by a factor of 125.
 
Looking at the bias obtained for $\fNL=1000$ shown in Figure~\ref{fig_bias_fNL_1000}, one immediately realizes that in contrast to the analytical formula of \citet{2005PhRvD..71j3514L} the bias extracted from numerical simulations does not scale quadratically with $\fNL$. This is particularly true for large angular scales. Here the agreement with the analytic expression, now phenomenologically amplified by a factor of 50, is much better than for $\fNL=100$. The fact that there is no simple scaling relation with $\fNL$ supports our concerns regarding the separability of lensing induced and primordial non-Gaussianity. To some extent the $\fNL$-dependence of the bias qualitatively approaches a quadratic form in the acoustic regime and beyond.

It is important to keep in mind that these are just qualitative considerations. The actual discrepancy between the bias obtained from a rigor simultaneous treatment of primordial as well as lensing induced non-Gaussianity via numerical simulations and its analytical lowest order estimate is tremendous. The analytical expression, even phenomenological amplified, is not suited for an accurate and consistent bias subtraction scheme. However, and this is the most important result, for intermediate values of $\fNL$, as suggested by current observational data, the bias associated with primordial non-Gaussianity of local type does not effect CMB lensing power reconstruction at all.

% --- Summary --- %
\section{Summary}
\label{sec_summary}

In this paper we revisited the influence of primordial non-Gaussianity on CMB lensing power reconstruction. 
We determined the bias associated with the non-trivial structure of the CMB trispectrum generated by non-Gaussian curvature perturbations of local-type.  While previous work of \citet{2005PhRvD..71j3514L} addressed this issue analytically, we used numerical simulations. This allows to overcome the non-obvious hypothesis that both effects leading to a connected part of the CMB four-point function do completely separate.
The results of this work can be summarized as follows:
 \begin{enumerate}
    \item The analytical expressions derived by \citet{2005PhRvD..71j3514L} in the limit of a flat sky have been extended 
    		to the all-sky formalism.
    \item In the case of non-Gaussian primordial curvature perturbations of local type the bias arising from a non-trivial trispectrum of the intrinsic CMB 
    		temperature contrast is substantially larger than so far expected.
    \item The limitation of the simulated non-Gaussian CMB maps (discussed in Section~\ref{subsec_numerical_sim}) does not change the conclusion. 
    		Discarding the physical mechanism and just thinking of the simulated maps as the simplest example of a scale dependent form 
		of non-Gaussian structure in the unlensed CMB, one finds that the associated connected part of the unlensed CMB is potentially 
		a severe contaminant in lensing power reconstruction studies. This interpretation of the simulated data addresses another important aspect. 
		In order to estimate the amplitude of the bias prior knowledge of the type of primordial non-Gaussianity is needed. For removing the bias this 
		prior knowledge is even more important.
    \item The strength of the contamination is obviously set by the amplitude of the non-Gaussian contributions. For current values of $\fNL$ only smallest  
    		multipoles are affected, while the effect on intermediate and small scales is completely negligible.  Low multipoles are anyway cosmic variance 
		dominated and have therefore only small predictive power. Thus, even in the full-sky case the bias arising from primordial non-Gaussianity 
		of  local type is not expected to compromise CMB lensing power reconstruction at all.
 \end{enumerate}
As this work focuses on the bias associated with primordial non-Gaussian curvature perturbations solely characterized by $\fNL$ an extension of this study including also the contributions from the $g_{\mathrm{NL}}$-term is desirable. However, the simulation of accurate CMB maps containing the correct non-Gaussian structure arising from both, $\fNL$ and $g_{\mathrm{NL}}$, is challenging.

% --- Acknowledgements --- %
\section*{Acknowledgements}

We would like to thank Franz Elsner and Benjamin D. Wandelt for making their non-Gaussian CMB simulations publicly available. PhMM acknowledges funding from the Graduate Academy Heidelberg and BMS's work is supported by the German Research Foundation (DFG) within the framework of the excellence initiative through the Heidelberg Graduate School of Fundamental Physics.

\bibliography{bibtex/aamnem,bibtex/references}
\bibliographystyle{mn2e}

\appendix

% --- appendix section: Some flat-sky expressions --- %
\section{Some flat-sky expressions}
\label{sec_flat_sky_expressions}

In this appendix we compile the flat-sky expressions used in this work. Starting with the lensing reconstruction bias linear in the lensing potential (cf. equation~\ref{eq_N_one}), the normalization reads
\begin{equation}
 A^{-1}(L) = \int\frac{\dd^2 l_1}{(2\pi)^2} \, f(\mathbf l_1, \mathbf l_2) g (\mathbf l_1, \mathbf l_2),
\end{equation}
while the optimal weights are given by
\begin{equation}
 g(\mathbf l_1, \mathbf l_2) = \frac{f(\mathbf l_1, \mathbf l_2)}{2 C^{\Theta\Theta}_{\ell_{1},\mathrm{obs}} C^{\Theta\Theta}_{\ell_{2},\mathrm{obs}}}
 						= \frac{\mathbf l_1 \cdot \mathbf L C^{\Theta\Theta}_{\ell_1}+\mathbf l_2 \cdot \mathbf L C^{\Theta\Theta}_{\ell_2}}{2 C^{\Theta\Theta}_{\ell_{1},\mathrm{obs}} C^{\Theta\Theta}_{\ell_{2},\mathrm{obs}}}
\end{equation}
with the additional condition $\mathbf l_1 + \mathbf l_2 = \mathbf L$ imposed.

Finally, the expression for the intermediate quantities $\mathcal{P}^{l_1l_2}_{l_3l_4}(L)$ which build up the flat-sky fully reduced CMB trispectrum associated with primordial non-Gaussianity of local type (cf. equation~\ref{eq_analytic_flat_sky_zeroth_bias}) reads
\begin{eqnarray}
 \mathcal P^{l_1l_2}_{l_3l_4}(L) &=&\int r_1^2\dd r_1\int r^2_2 \dd r_2  F_L(r_1,r_2) \nonumber\\
 && \phantom{\int}\times \left(\alpha_{\ell_1}(r_1) \beta_{\ell_2} (r_1)+ \alpha_{\ell_2}(r_1) \beta_{\ell_1} (r_1) \right) \nonumber \\
 && \phantom{\int}\times \left(\alpha_{\ell_3}(r_2) \beta_{\ell_4} (r_2)+ \alpha_{\ell_4}(r_2) \beta_{\ell_3} (r_2) \right)
\end{eqnarray}
with $F_L(r_1,r_2)$, $\alpha_{\ell}(r)$ and $\beta_\ell(r)$ defined in equations~\eqref{eq_cmb_trispectrum_F}~-~\eqref{eq_cmb_trispectrum_beta}.

\bsp

\label{lastpage}

\end{document}